\documentclass[12pt]{article}
\usepackage{amssymb,amsmath}

\newcommand{\half} {\frac{1}{2}}
\newcommand{\quart} {\frac{1}{4}}
\newcommand{\abs} [1] {\left|#1\right|}
\newcommand{\dash} {\nobreakdash-\hspace{0pt}}
\renewcommand{\eqref} [1] {(\ref{#1})}
\newcommand{\LW} {\Lambda_W}

\begin{document}

\begin{titlepage}
\hfill
\vbox{
    \halign{#\hfil         \cr
           } 
      }  
\vspace*{20mm}

\begin{center}
{\Large {\bf Black Holes in Ho\v{r}ava Gravity with\\ Higher Derivative Magnetic Terms}\\} \vspace*{15mm}

{\sc Eyal Gruss} \footnote{e-mail: {\tt eyalgruss@gmail.com}}

\vspace*{1cm}
{\it Raymond and Beverly Sackler School of Physics and Astronomy,\\
Tel-Aviv University, Tel-Aviv 69978, Israel.\\}

\end{center}

\vspace*{8mm}

\begin{abstract}
We consider Ho\v{r}ava gravity coupled to Maxwell and higher derivative magnetic terms. We construct static spherically symmetric black hole solutions in the low-energy approximation. We calculate the horizon locations and temperatures in the near-extremal limit, for asymptotically flat and (anti\dash)de Sitter spaces. We also construct a detailed balanced version of the theory, for which we find projectable and non-projectable, non-perturbative solutions.
\end{abstract}
\vskip 0.8cm

May 2010.

\end{titlepage}

\setcounter{footnote}{0}

\section{Introduction}

Ho\v{r}ava has recently proposed a theory of gravity which is power-counting renormalizable and possibly unitary \cite{horava}. This is achieved by assigning different scaling dimensions to time and space, at the cost of breaking Lorentz and diffeomorphism invariance at high energies. In four-dimensional Ho\v{r}ava gravity, one usually considers an anisotropic scaling exponent of $z=3$, such that $t\sim x^z$. This means that the Lagrangian should include higher spatial derivative terms up to sixth order. One may also consider couplings to scalar \cite{scalarvector1,scalar1,scalar2,scalarvector2} and vector \cite{scalarvector1,scalarvector2,magnetic,vector} fields with higher derivatives terms. There is an ongoing discussion regarding possible flaws of the theory. We refer to the latest reviews on these subjects \cite{problems}, and references therein.

The study of Black hole solutions and thermodynamics in general relativity with higher curvature corrections has been pioneered by \cite{CMP}. There, large black hole solutions were obtained by treating the higher curvature terms as a small perturbation. Black hole solutions in Ho\v{r}ava gravity and their thermodynamics have been discussed in \cite{deformed,charged,moresolutions,projectable,shift1,shift2,chargedprojectable,mass}. In particular, \cite{charged,chargedprojectable} discusses charged black holes in Ho\v{r}ava gravity coupled to the usual Maxwell term. In this work we add higher derivative magnetic terms to the mix, and discuss black hole solutions and thermodynamics.

The paper is organized as follows: In section $2$, we discuss Ho\v{r}ava-Maxwell theory with higher derivative magnetic terms. In section $3$, we discuss static spherically symmetric black hole solutions in the low-energy approximation. In section $3.1$, we calculate the horizon locations and temperatures for near-extremal asymptotically flat black holes. In section $3.2$, we do the same for near-cold asymptotically (anti\dash)de Sitter black holes. In section $4$, we discuss a detailed balanced version of the theory, and find non-perturbative solutions.

\section{Ho\v{r}ava-Maxwell Theory}

We use the ADM formalism where the metric is parameterized as
\begin{equation}
ds^2=-N^2dt^2+g_{ij}(dx^i+N^idt)(dx^j+N^jdt) \ ,
\end{equation}
where $i,j$ run over spatial coordinates. The extrinsic curvature is given by
\begin{equation}
K_{ij}=\frac{1}{2N}(\dot{g}_{ij}-\nabla_iN_j-\nabla_jN_i) \ , \qquad K\equiv K^i_{\phantom{i}i} \ ,
\end{equation}
where dot denotes a derivative with respect to $t$.
The four-dimensional Ricci scalar may be decomposed (see e.g. \cite{ricciadm}) as
\begin{equation}
\label{R4}
{}^{(4)}R=R+K_{ij}K^{ij}-K^2+(\textrm{covariant derivative}) \ ,
\end{equation}
where $R$ is the three-dimensional Ricci scalar. The last term contributes a total derivative in the Einstein-Hilbert Lagrangian, and may be dropped in this case. Note however, that this will not be true in general for higher curvature terms. Thus, Ho\v{r}ava gravity, which is of higher curvature in $R$, will differ from the more common ${}^{(4)}R$ higher curvature theories, even for static solutions. The Cotton tensor is given by
\begin{equation}
\label{cotton}
C_{ij}=\frac{\epsilon^{ikl}}{\sqrt{g}}\nabla_k\left(R^j_{\phantom{j}l}-\quart R\delta^j_l\right) \ ,
\end{equation}
where $\epsilon^{123}=1$, $g\equiv\mathrm{det}(g_{ij})$, and $R_{ij}$ is the three-dimensional Ricci tensor.

Ho\v{r}ava's ``detailed balance'' Lagrangian density \cite{horava} is given by: $\mathcal{L}_{Ho\check{r}ava}=\mathcal{L}_0+\mathcal{L}_1$,
with
\begin{eqnarray}
\label{Lhorava}
\frac{1}{N\sqrt{g}}\mathcal{L}_0&=&\frac{2}{\kappa^2}(K_{ij}K^{ij}-\lambda K^2)+\frac{\kappa^2\mu^2}{8(1-3\lambda)}(\LW R-3\LW^2) \ , \nonumber\\*
\frac{1}{N\sqrt{g}}\mathcal{L}_1&=&\frac{\kappa^2\mu^2(1-4\lambda)}{32(1-3\lambda)}R^2-\frac{\kappa^2\mu^2}{8}R_{ij}R^{ij}+\frac{\kappa^2\mu}{2w^2}\frac{\epsilon^{ijk}}{\sqrt{g}}R_{il}\nabla_jR^l_{\phantom{l}k}-\frac{\kappa^2}{2w^4}C_{ij}C^{ij} \ ,
\end{eqnarray}
where $\kappa$, $\lambda$, $\mu$, $\LW$ and $w$ are constant parameters. Comparing $\mathcal{L}_0$ to the Einstein-Hilbert Lagrangian, these parameters are related to the speed of light, Newton's constant and the cosmological constant by
\begin{equation}
c=\frac{\kappa^2\mu}{4}\sqrt{\frac{\LW}{1-3\lambda}} \ , \qquad G_N=\frac{\kappa^2}{32\pi c} \ , \qquad \Lambda=\frac{3}{2}\LW \ .
\end{equation}
We want this theory to reduce to general relativity in the low-energy limit. We therefore first set $\lambda=1$. Choosing units so that $c=1$, shows the coefficients of the $R^2$, $R_{ij}R^{ij}$ terms in $\mathcal{L}_1$ to be inversely proportional to ${\LW}$, with no additional free parameters. In order to allow both the cosmological constant term and the higher curvature terms to be small, we must deform the theory to break the relation between these terms \cite{horava,deformed}. As long as we do not change the form of the six-derivative $C_{ij}C^{ij}$ term, this is considered a soft violation of the detailed balance condition. We will write this as: $\mathcal{L}_{deformed}=\mathcal{L}_0+\epsilon a_0\LW\mathcal{L}_1$, where $a_0$ is a constant parameter, and $\epsilon>0$ will be used later as a small expansion parameter. In this parameterization $\LW$ factors out, leaving the now independent usual cosmological constant term (see e.g. \eqref{Lreduced}).

We now consider electromagnetic interactions. Maxwell's Lagrangian density in the ADM decomposition takes the form:
\begin{equation}
\frac{\gamma}{N\sqrt{g}}\mathcal{L}_2=\frac{2}{N^2}g^{ij}(F_{ti}-N^kF_{ki})(F_{tj}-N^lF_{lj})-F_{ij}F^{ij} \ ,
\end{equation}
where $\gamma=16\pi Gc^{-1}$ fixes the normalization in relation to the gravity part, and with the field strength given by
\begin{equation}
F_{ti}=\dot{A}_i-\partial_iA_t \ , \qquad F_{ij}=\partial_iA_j-\partial_jA_i \ ,
\end{equation}
where $(A_t,A_i)$ is the vector potential. We will add minimally coupled higher derivative magnetic terms as considered in \cite{scalarvector2,magnetic}:
\begin{equation}
\label{L3}
\frac{\gamma}{N\sqrt{g}}\mathcal{L}_3=-a_1\nabla_kF_{ij}\nabla^kF^{ij}-a_2\nabla_k\nabla_lF_{ij}\nabla^k\nabla^lF^{ij} \ ,
\end{equation}
where $a_1$, $a_2$ are constant parameters. \cite{magnetic} discusses constraints on the magnitude of $a_2$. Note that contrary to the case of \cite{scalarvector2,magnetic} which consider cosmological backgrounds, terms related by integration by parts to the RHS of \eqref{L3} are relevant since $N$ is space-dependant (in the non-projectable case). Nevertheless such terms are not considered here. Note also that in the spherically symmetric case considered later, $\nabla_jF^{ij}=0$ and $\triangle F^{ij}=\frac{1}{2g}\epsilon^{ijk}\epsilon^{lmn}R_{kl}F_{mn}$. Therefore, we have not considered here terms containing such derivatives. In section $4$, we consider a detailed balanced version of Ho\v{r}ava-Maxwell theory, which does include such terms.

We take our overall Lagrangian to be
\begin{equation}
\mathcal{L}=\mathcal{L}_0+\epsilon a_0\LW\mathcal{L}_1+\mathcal{L}_2+\epsilon\mathcal{L}_3 \ .
\end{equation}
Finally, we make the large black hole (or low-energy) approximation, in the spirit of \cite{CMP}. We require that the gravitational and magnetic higher derivative terms are small compared to $R$ and $F_{ij}F^{ij}$, respectively. In our analysis this will usually mean
\begin{equation}
\epsilon\abs{a_0}\ll r_{hor}^2 \ , \qquad \epsilon\abs{a_1}p^2\ll r_{hor}^4 \ , \qquad \epsilon\abs{a_2}p^2\ll r_{hor}^6 \ ,
\end{equation}
where $r_{hor}$ is the location of the horizon of interest, and $p$ is the magnetic charge. However, for asymptotically anti-de Sitter solutions with a large (negative) cosmological constant: $\abs{\LW}r_{hor}^2\gg1$, the conditions instead read
\begin{equation}
\epsilon\abs{a_0\LW}\ll1 \ , \qquad \epsilon\abs{a_1}p^2\ll r_{hor}^4 \ , \qquad \epsilon\abs{a_2\LW}p^2\ll r_{hor}^4 \ .
\end{equation}
This may be seen from \eqref{phi}. We will construct perturbative solutions of this theory to first order in $\epsilon$. Our calculations were done using Maple with GRTensor.

\section{Static Spherically Symmetric Black Holes}

We consider static spherically symmetric solutions, where the metric takes the form:
\begin{equation}
\label{gSSS}
ds^2=\Big(-N(r)^2+N_r(r)^2f(r)\Big)dt^2+2N_r(r)dtdr+\frac{1}{f(r)}dr^2+r^2(d\theta^2+sin^2\theta d\phi^2) \ ,
\end{equation}
where we have retained the shift function $N_r$ as discussed in \cite{projectable,shift1,shift2}. The vector potential takes the form:
\begin{equation}
\label{ASSS}
A_t=A(r) \ , \qquad A_r=A_\theta=0 \ , \qquad A_\phi=-p\cos\theta \ .
\end{equation}
Substituting the above ansatz into the Lagrangian density we get
\begin{eqnarray}
\label{Lreduced}
\frac{\sqrt{f}}{N}r^6\mathcal{L}_{reduced}&\propto&3\LW r^8+2f'r^7+2fr^6-2r^6+\epsilon a_0(2f'fr^5-2f'r^5-f^2r^4+2fr^4-r^4)+\nonumber\\*
&&{}-\frac{2f}{N^2}A'^2r^8+2p^2r^4+12\epsilon a_1p^2fr^2+3\epsilon a_2p^2(f'^2r^2-12f'fr+60f^2)+\nonumber\\*
&&{}+\frac{2f}{c^2N^2}(2N_r'N_rfr^7+N_r^2f'r^7+N_r^2fr^6) \ ,
\end{eqnarray}
where prime denotes a derivative with respect to $r$, and we have omitted an overall factor independent or $r$. Varying $\mathcal{L}_{reduced}$ with respect to $N$, $N_r$, $f$ and $A$, yields the equations of motion, the latter giving
\begin{equation}
A'=\frac{Nq}{\sqrt{f}r^2} \ ,
\end{equation}
where $q$ is the electric charge.

The variation by $N_r$ gives
\begin{equation}
\label{EOMNr}
\frac{N_rf^{3/2}r}{N}\left(\frac{2N'}{N}-\frac{f'}{f}\right)=0 \ .
\end{equation}
Let us first consider the case $N_r\neq0$. The motivation here is to obtain solutions to the projectable version of Ho\v{r}ava gravity, where $N=N(t)$ \cite{horava,projectable,chargedprojectable}. In our static case, $N$ must then be constant, implying that also $f$ is constant due to \eqref{EOMNr}. For $a_1=a_2=0$, uncharged and charged solutions of $N_r$ were discussed in \cite{projectable} and \cite{chargedprojectable}, respectively.\footnote{See e.g. \eqref{NrDB} with $\beta=0$, assuming $\epsilon a_0\LW=1$.} For general $a_1\neq0$ or $a_2\neq0$, we find there are no magnetically charged solutions to the equations of motion. \cite{shift2} discusses black hole thermodynamics when $N_r\neq0$. From here on we assume $N_r=0$.

The zeroth order solution is the Reissner-Nordstr\"om-(anti\dash)de Sitter black hole:
\begin{equation}
\label{N02}
N_0^2=f_0=-\frac{\LW}{2}r^2+1-\frac{2m_0}{r}+\frac{q^2+p^2}{r^2} \ ,
\end{equation}
where $m_0$ is the mass at zeroth order (see however \cite{mass}). For $a_1=a_2=0$, uncharged and charged non-perturbative solutions were obtained in \cite{deformed} and \cite{charged,chargedprojectable}, respectively.\footnote{See e.g. \eqref{N2DB} with $\beta=0$, assuming $\epsilon a_0\LW=1$.} For the general case, we will look for solutions to first order in $\epsilon$, in the form:
\begin{equation}
f=f_0(1+\epsilon\varphi) \ , \qquad N^2=f(1+\epsilon\eta) \ .
\end{equation}
Note that in the expression for $N^2$ we use $f$ rather than $f_0$, and truncate above order $\epsilon$. The solution reads
\begin{eqnarray}
\label{phi}
\varphi&=&\Bigg[105a_0\Big(\LW^2r^{12}-4\LW(q^2+p^2)r^8+16m_0^2r^6-16m_0(q^2+p^2)r^5+4(q^2+p^2)^2r^4\Big)+\nonumber\\*
&&{}+168a_1p^2\Big(15\LW r^8-10r^6+15m_0r^5-6(q^2+p^2)r^4\Big)+\nonumber\\*
&&{}+8a_2p^2\Big(-1575\LW^2r^8+2520\LW r^6-4095\LW m_0r^5+1764\LW(q^2+p^2)r^4-1890r^4+\nonumber\\*
&&{}\phantom{+8a_2p^2\Big(}+6930m_0r^3-6570m_0^2r^2-3240(q^2+p^2)r^2+6300m_0(q^2+p^2)r+\nonumber\\*
&&{}\phantom{+8a_2p^2\Big(}-1540(q^2+p^2)^2\Big)+\nonumber\\*
&&{}+1680m_0\delta r^9\Bigg]\Bigg/\Bigg[420\Big(\LW r^{12}-2r^{10}+4m_0r^9-2(q^2+p^2)r^8\Big)\Bigg] \ ,
\end{eqnarray}
and
\begin{equation}
\eta=3p^2\left(\frac{-a_1+9a_2\LW}{r^4}-\frac{10a_2}{r^6}+\frac{96a_2m_0}{7r^7}-\frac{4a_2(q^2+p^2)}{r^8}\right) \ ,
\end{equation}
where $\delta$ is an integration constant, and we have assumed the $\epsilon$-corrections vanish at infinity. Since in general $\eta\neq0$, $N^2$ is not proportional to $f$. Thus there is no transformation of coordinates that would yield a corresponding projectable solution \cite{projectable}, as indeed expected in light of the above discussion.

Looking at $N(r)^2$, one sees that the $\delta$-term is just an explicit $\epsilon$-correction to the mass: $m=m_0(1+\epsilon\delta)$. We assume the horizon locations of the black hole are given by the roots of $g_{tt}=N(r)^2=0$, as in the relativistic case (see \cite{shift1} for a discussion of the notion of horizon in Ho\v{r}ava gravity). One may set $\delta$ to make $\varphi(r)$ and $\eta(r)$ nonsingular on the unperturbated black hole event horizon, as done in \cite{CMP}. This would mean the event horizon location does not get an $\epsilon$-correction. However, the mass then has a fixed $\epsilon$-correction, and this depends nontrivially on the zeroth order mass $m_0$. Also, other horizons of the solution do get $\epsilon$-corrections. We will take the alternate route of leaving $\delta$ free, and finding all horizon locations perturbatively, designating: $r_{hor}=r_0(1+\epsilon\rho)$, with $\epsilon\abs{\rho}\ll1$. Note that in the near-horizon region, $f(r)$ and $N^2(r)$ now get an additional $\epsilon$-correction due to $\rho$ in $f_0(r)$, and it is the overall correction that should be small.

The Hawking temperature of our static spherically symmetric black hole (with $N_r=0$, otherwise see \cite{shift2}) is given by
\begin{equation}
T=\frac{\abs{g_{tt}'}}{4\pi\sqrt{-g_{tt}g_{rr}}}\Bigg|_{r=r_{hor}}=\frac{\abs{N'}\sqrt{\abs{f}}}{2\pi}\Bigg|_{r=r_{hor}}
\ .
\end{equation}
Note that for asymptotically de Sitter black holes, the temperature may need to include the Bousso-Hawking normalization factor \cite{dstemp}, not considered here. We do not calculate the Bekenstein-Hawking area law entropy, as it is expected that the entropy would be corrected by the higher derivative terms. In fully diffeomorphism invariant theories, this is computed by the Wald entropy \cite{wald}. However, as mentioned also in \cite{shift1}, it is not clear how to do this in our diffeomorphism breaking theory. In the following we will analyze the horizon locations and temperatures for some simplified cases of black holes.

\subsection{Asymptotically Flat Black Holes}
We first consider asymptotically flat black holes with $\LW=0$ (holding $c$ finite). The zeroth order inner and outer horizons are given by
\begin{equation}
r_0^{in}=m_0-\sqrt{m_0^2-q^2-p^2} \ , \qquad r_0^{out}=m_0+\sqrt{m_0^2-q^2-p^2} \ .
\end{equation}
The expressions for the first order corrections are lengthy, and to be concise we will consider only the extremal and near-extremal cases. The extremal limit is determined by: $r^{in}=r^{out}$, which is satisfied at zeroth order by
\begin{equation}
r_0^{extr}=m_0^{extr}=\sqrt{q^2+p^2} \ .
\end{equation}
At this extremum point, the first order correction to the horizon location will not affect $N(r)^2$ to first order in $\epsilon$. In order to satisfy $N(r)^2=0$, one must set
\begin{equation}
\delta^{extr}=\frac{-105a_0(q^2+p^2)^2+42a_1(q^2+p^2)p^2+20a_2p^2}{420(q^2+p^2)^3} \ ,
\end{equation}
correcting the extremal mass. It is conjectured that the mass-charge ratio of extremal black holes is decreased by higher curvature corrections \cite{mass_charge}, implying: $\delta^{extr}<0$. The first order correction to the extremal horizon may then be got either by taking the limit of the non-extremal case, or by satisfying: $\left(N(r)^2\right)'=0$, that is, having the temperature vanish. It reads
\begin{equation}
\rho^{extr}=-\frac{a_0}{4(q^2+p^2)} \ .
\end{equation}
Note that the extremal horizon does not depend on $a_1$, $a_2$.

In the near-extremal approximation we parameterize the mass as
\begin{equation}
m=m^{extr}\sqrt{1+\Delta^2} \ , \qquad m^{extr}=m_0^{extr}(1+\epsilon\delta^{extr}) \ ,
\end{equation}
and expand all expressions to leading order in $\Delta\ll1$. The horizon locations satisfying: $N(r)^2=0+O(\Delta^3)$, are given at zeroth order in $\epsilon$ by
\begin{equation}
r_0^{in}=\sqrt{q^2+p^2}(1-\Delta)+O(\Delta^2) \ , \qquad r_0^{out}=\sqrt{q^2+p^2}
(1+\Delta)+O(\Delta^2) \ ,
\end{equation}
and at first order in $\epsilon$ by
\begin{eqnarray}
\rho^{in}&=&\frac{-105a_0(q^2+p^2)^2-336a_1(q^2+p^2)p^2\Delta-100a_2p^2\Delta}{420(q^2+p^2)^3}+O(\Delta^2) \ , \nonumber\\ \rho^{out}&=&\frac{-105a_0(q^2+p^2)^2+336a_1(q^2+p^2)p^2\Delta+100a_2p^2\Delta}{420(q^2+p^2)^3}+O(\Delta^2) \ .
\end{eqnarray}
The Hawking temperature is
\begin{equation}
T^{out}=\frac{\Delta}{2\pi\sqrt{q^2+p^2}}\left(1+\epsilon\frac{
525a_0(q^2+p^2)^2-294a_1(q^2+p^2)p^2-80a_2p^2}{420(q^2+p^2)^3}\right)+O(\Delta^2+\epsilon^2) \ .
\end{equation}

\subsection{Asymptotically (Anti\dash)de Sitter Black Holes}
We now consider asymptotically (anti\dash)de Sitter black holes with $\LW\neq0$. In the de Sitter case, having $\LW>0$ (holding $c$ positive by analytic continuation), there exists a cosmological horizon denoted: $r^{cosm}$. Let us recall the classification of the limiting cases of Reissner-Nordstr\"om-de Sitter black holes \cite{rnds,dstemp}:
\begin{enumerate}
\item[(a)] lukewarm: $r^{in}<r^{out}<r^{cosm}$ and $T^{out}=T^{cosm}$.
\item[(b)] charged Nariai: $r^{in}<r^{out}=r^{cosm}$.
\item[(c)] cold: $r^{in}=r^{out}<r^{cosm}$.
\item[(d)] ultracold: $r^{in}=r^{out}=r^{cosm}$.
\end{enumerate}
In the anti-de Sitter case, $\LW<0$, and the only relevant limit is the cold black hole: $r^{in}=r^{out}$. Together with the flat case, these limiting cases (and their ``near'' versions) have a concise zeroth order form for the horizon location. We will first consider the cold and ultracold limits. We will then extend to the near-cold black hole: $r^{in}\lesssim r^{out}$. For the latter to be concise also at first order in $\epsilon$, we will consider the near-flat approximation: $\abs{\LW}(q^2+p^2)\ll1$.

The zeroth order cold black hole metric function is of fourth degree in $r$, and has a double zero. It may be written as
\begin{equation}
\label{doublezero}
\left(N_0^{cold}\right)^2=-\frac{\LW}{2r^2}(r-A)(r-B)^2(r-C) \ ,
\end{equation}
where $A$, $B$, $C$ are constants. Equating \eqref{doublezero} with \eqref{N02} and comparing powers of $r$, one gets
\begin{eqnarray}
A+2B+C&=&0 \ , \nonumber\\
3\LW B^2+2\LW BC+\LW C^2-2&=&0 \ , \nonumber\\
\LW B^3-B+m_0&=&0 \ , \nonumber\\
3\LW B^4-2B^2+2(q^2+p^2)&=&0 \ .
\end{eqnarray}
This gives
\begin{equation}
\label{r0cold}
r_0^{cold}=B=\sqrt{\frac{2(q^2+p^2)}{\sqrt{1-6\LW(q^2+p^2)}+1}} \ ,
\end{equation}
and
\begin{equation}
m_0^{cold}=\frac{B}{3}\left(2+\sqrt{1-6\LW(q^2+p^2)}\right) \ ,
\end{equation}
where for $\LW>0$ we have assumed $B\leq C$, and in that case we have also
\begin{equation}
r_0^{cosm}=C=-B+\sqrt{\frac{4+2\sqrt{1-6\LW(q^2+p^2)}}{3\LW}} \ .
\end{equation}
As in the extremal flat case, to satisfy $N(r)^2=0$ to first order in $\epsilon$, one must correct the cold black hole mass by setting
\begin{eqnarray}
\delta^{cold}&=&\Bigg[70a_0(q^2+p^2)^2\Big(-7+12\LW(q^2+p^2)-2\sqrt{1-6\LW(q^2+p^2)}\Big)+\nonumber\\*
&&{}+126a_1(q^2+p^2)p^2\Big(1-18\LW(q^2+p^2)+\sqrt{1-6\LW(q^2+p^2)}\Big)+\nonumber\\*
&&{}+3a_2p^2\Big(20-318\LW(q^2+p^2)+2844\LW^2(q^2+p^2)^2-23\sqrt{1-6\LW(q^2+p^2)}+\nonumber\\*
&&{}\phantom{+3a_2p^2\Big(}+43\left(1-6\LW(q^2+p^2)\right)^{3/2}\Big)\Bigg]\frac{1+\sqrt{1-6\LW(q^2+p^2)}}{1680(q^2+p^2)^3\left(2+\sqrt{1-6\LW(q^2+p^2)}\right)} \ , \nonumber\\*
\end{eqnarray}
The first order correction to the cold horizon location reads
\begin{equation}
\label{rhocold}
\rho^{cold}=-a_0\frac{1+\sqrt{1-6\LW(q^2+p^2)}}{8(q^2+p^2)\sqrt{(1-6\LW(q^2+p^2)}} \ .
\end{equation}
Here again the cold horizon does not depend on $a_1$, $a_2$. The full expressions for the cosmological horizon location and temperature are more cumbersome, and will be given later in the near-flat limit.

The ultracold limit is determined by: $r^{cold}=r^{cosm}$, which is satisfied at zeroth order by
\begin{equation}
\label{UC}
\Lambda_0^{UC}=\frac{1}{6(q^2+p^2)} \ ,
\end{equation}
and
\begin{equation}
r_0^{UC}=\frac{3}{2}m_0^{UC}=\sqrt{2(q^2+p^2)} \ ,
\end{equation}
where we have denoted: $\LW=\Lambda_0(1+\epsilon\upsilon)$, and we expand in $\epsilon$ before taking the ultracold limit. Note that the first order corrections to the mass and horizon location, now also get a contribution from their respective zeroth order terms due to $\upsilon$. Thus the $\epsilon$-correction in the expansion of \eqref{r0cold}, serves to cancel out the divergency of \eqref{rhocold} in the limit \eqref{UC}. The overall first order terms are given by
\begin{eqnarray}
\upsilon^{UC}&=&\frac{a_0}{2(q^2+p^2)} \ , \nonumber\\
\delta^{UC}&=&\frac{-280a_0(q^2+p^2)^2-126a_1(q^2+p^2)p^2+69a_2p^2}{1680(q^2+p^2)^3} \ , \nonumber\\
\rho^{UC}&=&-\frac{a_0}{4(q^2+p^2)} \ .
\end{eqnarray}

In the near-cold approximation we parameterize the mass as
\begin{equation}
m=m^{cold}\sqrt{1+\Delta^2} \ , \qquad m^{cold}=m_0^{cold}(1+\epsilon\delta^{cold}) \ ,
\end{equation}
and expand all expressions to leading order in $\Delta\ll1$. The horizon locations satisfying: $N(r)^2=0+O(\Delta^3)$, are given at zeroth order in $\epsilon$ by
\begin{equation}
r_0^{in}=b(1-\alpha_b\Delta)+O(\Delta^2) \ , \qquad r_0^{out}=b
(1+\alpha_b\Delta)+O(\Delta^2) \ , \qquad r_0^{cosm}=c
(1+\alpha_c\Delta^2)+O(\Delta^3) \ ,
\end{equation}
where
\begin{eqnarray}
\alpha_b&=&\sqrt{\frac{1}{3}+\frac{2}{3\sqrt{1-6\LW(q^2+p^2)}}} \ , \nonumber\\
\alpha_c&=&\frac{(2+S)\left(1-S-\sqrt{4-2S-2S^2}\right)}{6\left(8-2S^2-4\sqrt{4-2S-2S^2}-S\sqrt{4-2S-2S^2}\right)} \ , \qquad S=\sqrt{1-6\LW(q^2+p^2)}\ . \nonumber\\*
\end{eqnarray}

We now take the near-flat approximation by expanding to leading order in $L=\abs{\LW}(q^2+p^2)\ll1$. The mass, the horizon locations and the temperatures of the near-cold near-flat black hole are given by
\begin{eqnarray}
m_0^{cold}&=&\sqrt{q^2+p^2}\left(1-\frac{L}{4}\right)+O(L^2) \ , \nonumber\\
r_0^{in}&=&\sqrt{q^2+p^2}\left(1+\frac{3L}{4}-\Delta-\frac{7L\Delta}{4}\right)+O(L^2+\Delta^2) \ , \nonumber\\
r_0^{out}&=&\sqrt{q^2+p^2}\left(1+\frac{3L}{4}+\Delta+\frac{7L\Delta}{4}\right)+O(L^2+\Delta^2) \ , \nonumber\\
r_0^{cosm}&=&\sqrt{q^2+p^2}\left(\sqrt{2}L^{-1/2}-1-\frac{\Delta^2}{2}\right)+O(L^{1/2}+\Delta^4) \ , \nonumber\\
\delta^{cold}&=&\delta^{extr}_{L=0}+\frac{525a_0(q^2+p^2)^2-924a_1(q^2+p^2)p^2-656a_2p^2}{840(q^2+p^2)^3}L+O(L^2) \ , \nonumber\\
\rho^{in}&=&\rho^{in}_{L=0}+\frac{105a_0(q^2+p^2)^2(-3+4\Delta)+1680a_1(q^2+p^2)p^2\Delta+1748a_2p^2\Delta}{840(q^2+p^2)^3}L+O(L^2+\Delta^2) \ , \nonumber\\
\rho^{out}&=&\rho^{out}_{L=0}-\frac{105a_0(q^2+p^2)^2(3+4\Delta)+1680a_1(q^2+p^2)p^2\Delta+1748a_2p^2\Delta}{840(q^2+p^2)^3}L+O(L^2+\Delta^2) \ , \nonumber\\
\rho^{cosm}&=&-\frac{\delta^{extr}_{L=0}L^{1/2}}{\sqrt{2}}\left(1+\frac{\Delta^2}{2}\right)+O(L+\Delta^4) \ , \nonumber\\
T^{out}&=&T^{out}_{L=0}-\frac{11L\Delta}{8\pi\sqrt{q^2+p^2}}\left(1+\epsilon\frac{5985a_0(q^2+p^2)^2-6090a_1(q^2+p^2)p^2-3464a_2p^2}{4620(q^2+p^2)^3}\right)+\nonumber\\
&&{}+O(L^2+\Delta^2+\epsilon^2) \ , \nonumber\\
T^{cosm}&=&\frac{\sqrt{2}L^{1/2}-2L-L\Delta^2}{4\pi\sqrt{q^2+p^2}}\left(1-\sqrt{2}\epsilon\delta^{extr}_{L=0}L^{1/2}\left(1+\frac{\Delta^2}{2}\right)\right)+O(L^{3/2}+\Delta^4+\epsilon^2) \ .
\end{eqnarray}
where $L=0$ designates the near-extremal flat case results of the previous section. For $\Delta=0$, we get the cold near-flat limit.

\section{A Detailed Balanced Ho\v{r}ava-Maxwell Theory}

Detailed balanced Ho\v{r}ava gravity was presented in \cite{horava}. A detailed balanced version of the theory coupled to a scalar field, was derived in \cite{scalar2}. Here we discuss a detailed balanced version of Ho\v{r}ava gravity coupled to Maxwell and higher derivative magnetic terms. Detailed balance is a symmetry condition which serves to reduce the number of terms in the Lagrangian. This allows in our case to yield analytic non-perturbative solutions.

The detailed balance Lagrangian density will be given by: $\mathcal{L}=\mathcal{L}_K-\mathcal{L}_V$. The kinetic part is the same as before:
\begin{equation}
\frac{1}{N\sqrt{g}}\mathcal{L}_K=\frac{2}{\kappa^2}(K_{ij}K^{ij}-\lambda K^2)+\frac{2}{\gamma N^2}g^{ij}(F_{ti}-N^kF_{ki})(F_{tj}-N^lF_{lj}) \ .
\end{equation}
The potential part is defined to be of the form:
\begin{equation}
\frac{1}{N\sqrt{g}}\mathcal{L}_V=\frac{\kappa^2}{8}\mathcal{G}_{ijkl}E^{ij}E^{kl}+\frac{1}{2\gamma}E_iE^i \ ,
\end{equation}
with the generalized de Witt metric:
\begin{equation}
\mathcal{G}_{ijkl}=\half(g_{ik}g_{jl}+g_{il}g_{jk})+\frac{\lambda}{1-3\lambda}g_{ij}g_{kl} \ ,
\end{equation}
and where
\begin{equation}
E^{ij}=\frac{1}{\sqrt{g}}\frac{\delta W}{\delta g_{ij}} \ , \qquad E^i=\frac{1}{\sqrt{g}}\frac{\delta W}{\delta A_i} \ ,
\end{equation}
for some three-dimensional Euclidean Lagrangian density $W$. For $z=3$, $W$ should include up to three spatial derivatives.

The gravitational part of $W$, due to \cite{horava}, is given by
\begin{equation}
\frac{1}{\sqrt{g}}W_{Ho\check{r}ava}=-\frac{1}{w^2}\frac{\epsilon^{ijk}}{\sqrt{g}}\Gamma^{m}_{\phantom{m}il}\left(\partial_j\Gamma^l_{\phantom{l}km}+\frac{2}{3}\Gamma^l_{\phantom{l}jn}
\Gamma^{n}_{\phantom{n}km}\right)+\mu(R-2\LW) \ ,
\end{equation}
where $\Gamma^{i}_{\phantom{i}jk}$ are the Christoffel symbols. This is the Lagrangian density of topological massive gravity \cite{tmg}, with the first term being the gravitational Chern-Simons term\footnote{Note that to be consistent with \eqref{Eijhor} and \eqref{cotton}, we find this term to have an opposite sign compared to \cite{horava}.}. Varying with respect to $g_{ij}$ yields
\begin{equation}
\label{Eijhor}
E^{ij}_{Ho\check{r}ava}=\frac{1}{\sqrt{g}}\frac{\delta W_{Ho\check{r}ava}}{\delta g_{ij}}=\frac{2}{w^2}C^{ij}-\mu\left(R^{ij}-\half Rg^{ij}+\LW g^{ij}\right) \ .
\end{equation}
Using this, one gets the potential part in $\mathcal{L}_{Ho\check{r}ava}$ \eqref{Lhorava}.

For the electromagnetic part of $W$, we take
\begin{equation}
\frac{1}{\sqrt{g}}W_{EM}=\frac{\epsilon^{ijk}}{\sqrt{g}}A_i\partial_jA_k+b_1F_{ij}F^{ij}+b_2\frac{\epsilon^{ijk}}{\sqrt{g}}(\triangle A_i-R^l_{\phantom{l}i}A_l)\partial_jA_k \ ,
\end{equation}
where $b_1$, $b_2$ are constant parameters. This is the three-dimensional Chern-Simons-Maxwell-extended-Chern-Simons Lagrangian density in curved space \cite{csmecs}. Varying with respect to $g_{ij}$ yields
\begin{eqnarray}
E^{ij}_{EM}=\frac{1}{\sqrt{g}}\frac{\delta W_{EM}}{\delta g_{ij}}&=&b_1\left(\half F_{kl}F^{kl}g^{ij}-2g_{kl}F^{ki}F^{lj}\right)+\nonumber\\* &&{}+b_2\frac{\epsilon^{ikl}}{\sqrt{g}}(F^{jm}\nabla_kF_{lm}+F_{lm}\nabla_kF^{jm}-g_{kn}F_{lm}\nabla^jF^{mn}) \ ,
\end{eqnarray}
where in the spherically symmetric case, the $b_2$ term vanishes. Varying with respect to the $A_i$ yields
\begin{equation}
E^i=\frac{1}{\sqrt{g}}\frac{\delta W_{EM}}{\delta A_i}=\frac{\epsilon^{ijk}}{\sqrt{g}}F_{jk}+4b_1\nabla_jF^{ij}+b_2\frac{\epsilon^{jkl}}{\sqrt{g}}\left(\delta^i_j\triangle F_{kl}-R^i_{\phantom{i}j}F_{kl}\right) \ ,
\end{equation}
where in the spherically symmetric case, the $b_1$ and $b_2$ terms vanish. Note that both variations yield gauge invariant expressions. Other terms may be added to $W_{EM}$, which would yield gauge dependant expressions. Nevertheless, some of these may still lead to gauge invariant equations of motion.

In the static spherically symmetric case, we substitute the ansatz \eqref{gSSS}, \eqref{ASSS} into the Lagrangian density to get
\begin{eqnarray}
\frac{\sqrt{f}}{N}r^6\mathcal{L}_{reduced}&\propto&3\LW r^8+2f'r^7+2fr^6-2r^6+\frac{1}{\LW}(2f'fr^5-2f'r^5-f^2r^4+2fr^4-r^4)+\nonumber\\*
&&{}-\frac{2f}{N^2}A'^2r^8+2p^2r^4+\frac{\beta p^2}{\LW}(2\LW r^4-2f'r^3+6fr^2-6r^2-5\beta p^2)+\nonumber\\*
&&{}+\frac{2f}{c^2N^2}(2N_r'N_rfr^7+N_r^2f'r^7+N_r^2fr^6) \ ,
\end{eqnarray}
where $\beta\equiv\mu^{-1}b_1$, we have set $\lambda=1$, and have omitted an overall factor independent or $r$.

We now find non-perturbative solutions with higher order magnetic terms. For the projectable case with $N_r\neq0$, $N=1$, and constant $f$, we find the solution:
\begin{equation}
\label{NrDB}
N_r=\pm\frac{c}{f}\sqrt{\frac{\LW}{2}r^2+f-1+\frac{2m}{r}+\frac{(f-1)^2}{2\LW r^2}-\frac{q^2+p^2}{r^2}+\beta p^2\left(-\frac{1}{r^2}+\frac{1-f}{\LW r^4}+\frac{\beta p^2}{2\LW r^6}\right)} \ .
\end{equation}
For the non-projectable case with $N_r=0$, we find
\begin{equation}
\label{N2DB}
N^2=f=-\LW r^2+1+\beta\frac{p^2}{r^2}-\sqrt{-4\LW mr+2\LW(q^2+p^2)} \ .
\end{equation}
These generalize the solutions with $\beta=0$ of \cite{charged,chargedprojectable}.

\section*{Acknowledgements}

I would like to thank Yaron Oz for his support and guidance.

\end{document}